\documentstyle[11pt,newpasp,twoside,epsf]{article}
\reversemarginpar

\begin{document}

\title{The Highly Relativistic Binary Pulsar PSR J0737-3039A: 
Discovery and Implications}

\author{M.~Burgay$^1$, N.~D'Amico$^{2,1}$, A.~Possenti$^1$, 
R.N.~Manchester$^3$, A.G.~Lyne$^4$, B.C.~Joshi$^5$, 
M.A.~McLaughlin$^4$, M.~Kramer$^4$, J.M.~Sarkissian$^6$, 
F.~Camilo$^7$, V.~Kalogera$^8$, C.~Kim$^8$, D.R.~Lorimer$^4$}
\affil{$^1$INAF -Osservatorio Astronomico di Cagliari, Loc. 
           Poggio dei Pini, Strada 54, 09012 Capoterra, Italy, 
       $^2$Universit\`a degli Studi di Cagliari, Dipartimento di 
           Fisica, SP Monserrato-Sestu km 0.7, 09042 Monserrato, Italy,
       $^3$Australia Telescope National Facility, CSIRO, P.O. Box
	   76, Epping, New South Wales 2121, Australia, 
       $^4$University of Manchester, Jodrell Bank Observatory, 
           Macclesfield, Cheshire, SK11 9DL, UK,
       $^5$National Center for Radio Astrophysics, P.O. Bag 3, 
           Ganeshkhind, Pune 411007, India, 
       $^6$Australia Telescope National Facility, CSIRO, Parkes 
           Observatory, P.O. Box 276, Parkes, New South Wales
	   2870, Australia,        
       $^7$Columbia Astrophysics Laboratory, Columbia
	   University, 550 West 120 th Street, New York, New York 10027,
       $^8$Northwestern University, Department of Physics and
	   Astronomy, Evanston, Illinois 60208, USA}

\begin{abstract}
PSR~J0737$-$3039A is a millisecond pulsar with a spin period of 22.7 ms
included in a double-neutron star system with an orbital period of 2.4
hrs. Its companion has also been detected as a radio pulsar, making
this binary the first known double-pulsar system. Its discovery has
important implications for relativistic gravity tests, gravitational
wave detection and plasma physics. Here we will shortly describe the
discovery of the first pulsar in this unique system and present the
first results obtained by follow-up studies.
\end{abstract}

\section{Introduction}
Since the discovery of the first binary pulsar (Hulse \& Taylor 1975),
the detection of two active pulsars in the same binary
system has been a primary aim of any pulsar survey.  We here summarize
the basic steps which eventually led to the discovery of this
long-sought system and report on the first implications for
gravitational waves detection.  Papers by Kramer et al. and Manchester
et al. (in these proceedings) will give more details on the second
discovered pulsar, dealing with the opportunity to use this binary as
a magnificent laboratory of relativistic gravity and for investigating
magnetospheric processes.

\section{The discovery}

PSR~J0737$-$3039A, a millisecond pulsar with a spin period of $22.7$ ms,
was discovered in April 2003 (Burgay et al. 2003) in a 4-minute
pointing of the Parkes High-Latitude Pulsar Survey (Burgay et al. in
preparation). The original detection plot is shown in Figure 1,
reporting a signal-to-noise ratio of 18.7. A higher signal-to-noise
ratio of 26.3 resulted from the same observation for the first
harmonics of the pulse period ($\sim 11.3$ ms), due to the similar
energy contribution and roughly half-phase separation of the two peaks
of the profile displayed in Figure 1; the pulse period ambiguity was
easily solved as soon as further longer observations became available
(a high resolution pulse profile for PSR~J0737$-$3039A is presented in
Manchester et al. 2004, these proceedings).  

\begin{figure}
\begin{center}
\plotfiddle{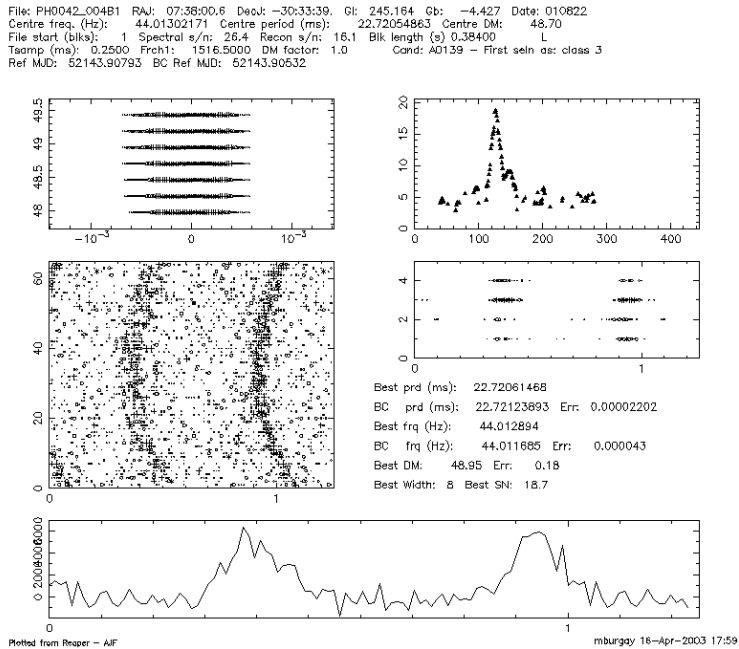}{9truecm}{270}{50}{50}{-215}{280}
\caption{Original detection plot for PSR~J0737$-$3039A. The pulsar was
detected in the 11$^{\rm th}$ beam (beam B) of the Multibeam receiver
of the Parkes 64 meter radiotelescope with a signal-to-noise ratio of
18.7. The offset between the sky coordinates of the centre of the beam
and the (subsequently determined) position of PSR~J0737$-$3039A is 
$6\farcm3.$}
\end{center}
\end{figure} 

From the very pronounced curvature in the spin phase vs
sub-integrations box (central panel on the left of the Figure 1),
denoting a time depending change in the phase of arrival of the
pulses, it was immediately clear that the pulsar signal was affected
by a significant Doppler effect. That in turn suggested that the
pulsar was experiencing the gravitational pull of a nearby and
relatively massive companion.

Using a code suitable for searching strongly accelerated pulsars, we
found the best correction to the Doppler phase shift for an
acceleration of 99 m/s$^2$ suggesting a binary period of just few
hours for a companion of $\sim 1~{\rm M_\odot}$. Follow-up observations
performed in May 2004, consisting of three $\sim 5$-hour integrations,
confirmed that the orbit is indeed very tight and far from being
circular: the binary period $P_b$ is only 2.4 hr and the eccentricity
$e \sim 0.09$. This makes J0737$-$3039A's orbit the tightest among
those of all known binary pulsars in eccentric systems. Figure 2 shows
the radial velocity curve obtained plotting the barycentric spin
period of the pulsar, measured at different times, versus the binary
phase: the fact that the orbit is eccentric is easily seen from
the asymmetric shape of the curve.

\begin{figure}[htbp]
\begin{center}
\plotfiddle{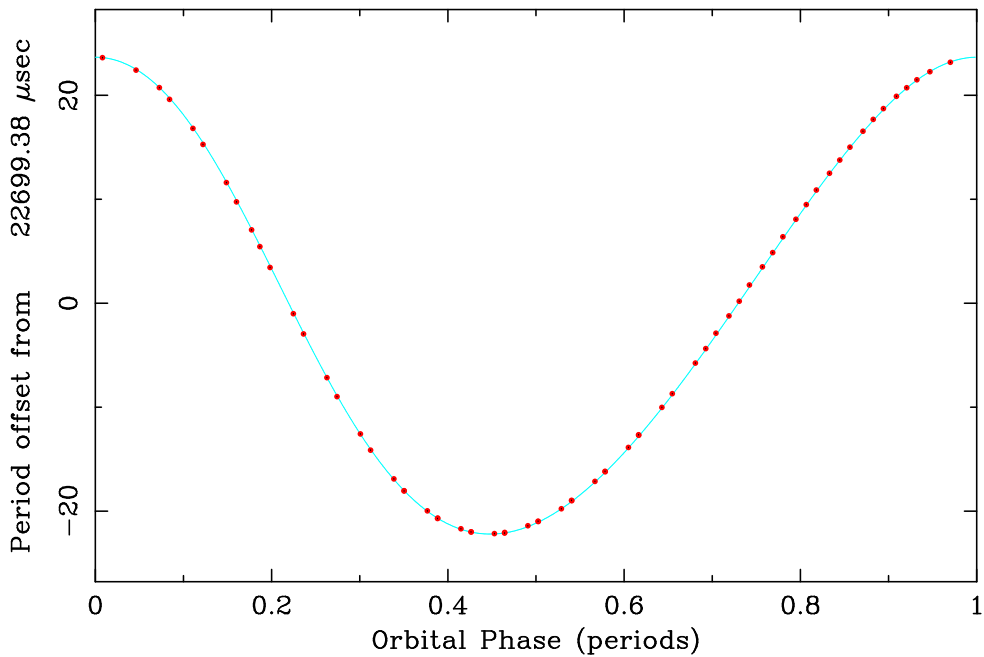}{9truecm}{270}{55}{55}{-185}{280}
\caption{The original radial velocity plot for PSR~J0737$-$3039A,
obtained from follow-up observations taken at Parkes early in May
2003.}
\end{center}
\end{figure}

From the approximate orbital parameters available after two days of
follow-up observations ($P_b \sim 2.4$ hr,$a \sin{i} \sim 1.4$ lt-s;
for the current best estimates of these parameters see Table 1 in
Kramer et al., these proceedings), the pulsar mass function was
calculated and resulted $M_f = 0.29$ M$_\odot,$ implying a minimum
companion mass of about 1.24 M$_\odot$, assuming $M_{NS} =
1.35$M$_\odot$. According to this mass, the companion star could be a
non degenerate object, a massive Carbon-Oxygen white dwarf (CO-WD) or
a second neutron star.  The first hypothesis was immediately ruled out
since the radius of a non degenerate object of the required mass would
almost completely fill the orbit of the system, probably strongly
affecting the radio emission, which is not seen in the signal of
PSR~J0737$-$3039A.  On the other hand, also the CO-WD scenario
appeared unlikely considering the eccentricity of the orbit: a binary
containing a recycled neutron star and a white dwarf is indeed
expected to be highly circular since it had the time to reach the
minimum energy configuration (i.e. to circularise). If a second,
recent supernova explosion has occurred forming another neutron star,
the energy and momentum released can have distorted the system and
this would explain the observed eccentricity. 

A further strong constraint on the nature of PSR~J0737$-$3039A
companion came few days later from the first fit of a binary model to
the times of arrival of the pulsations.  By using a data span of only
6 days, a 10-$\sigma$ determination of the advance of the periastron,
$\dot{\omega}$, was possible. This parameter resulted to have a
remarkably high value: $\dot{\omega}\sim 17^\circ/$yr (note that the
the previously highest observed value was $5.33^\circ/$yr, for
PSR~J1141$-$6545, Kaspi et al. 2000).  If interpreted in the framework
of general relativity, the measured value of $\dot{\omega}$ implied a
total mass for the system containing PSR~J0737$-$3039A of about 2.58
M$_\odot$ giving a maximum mass for the pulsar of about 1.34 M$_\odot$
and a minimum mass for the companion $\sim 1.24$ M$_\odot$. While the
maximum mass for the pulsar perfectly agrees with the other
measurements of neutron star masses (Thorsett \& Chakrabarty 1999),
the mass of the companion is a little lower than average. In absence
of additional information the white dwarf hypothesis could not be
completely rejected. It is important to point out, anyway, that the
$\dot{\omega}$ value that one measures, in general, is given by the
term of equation 1 plus two classical extra terms, arising {\it (i)}
from tidal deformations of the companion star (relevant only if the
companion is non degenerate) and {\it (ii)} from rotationally induced
quadrupole moment of the companion star, applicable to the case of a
fast rotating white dwarf. For a neutron star both the additional
contributions are negligible.  If the companion to J0737$-$3039A was a
white dwarf the relativistic $\dot{\omega}$, and by consequence the
total system mass, would be smaller than the measured one
(since $\dot\omega_{GR} = \dot\omega_{obs} - \dot\omega_{classical}$)
implying an implausibly small ($\la 1$ M$_\odot$) maximum allowed mass
for J0737$-$3039A (that, being a pulsar, is certainly a neutron
star). All these pieces of evidence strongly suggested that the
discovered binary was the sixth, and by far the most relativistic,
Double Neutron Star (DNS) system known. The ultimate confirmation of
the above picture came few months later when, analysing the follow-up
observations of PSR~J0737$-$3039A, a strong signal with a repetition
period of $\sim$ 2.8 seconds occasionally appeared (Lyne at
al. 2004). The newly discovered pulsar, henceforth called
PSR~J0737$-$3039B (or simply `B'), had the same dispersion measure as
PSR~J0737$-$3039A (or `A'), and showed orbital Doppler variations that
identified it, without any doubt, as the companion to the millisecond
pulsar. The first ever Double Pulsar system had been eventually
discovered.

\section{Determination of Post-Keplerian parameters} 
In the binary system containing PSR~J0737$-$3039A and B, the
relativistic effects are highly enhanced, thanks to its short orbital
period and to its high orbital inclination. That makes it an excellent
laboratory for studying relativistic gravity. Using only the follow up
observations of pulsar A, we have been able, in less than a year of
regular timing, to measure four Post-Keplerian parameters, whose
values are listed in Table 1 of Kramer et al. (these proceedings):
besides the already mentioned parameter $\dot{\omega}$, we have
measured $\gamma$ (namely the term taking into account gravitational
redshift and time dilation) and the parameters $r$ and $s\equiv
\sin{i}$. The latter represent respectively the range and the shape of
the Shapiro delay, measuring the time delays of the signal caused by
the space-time deformations around the companion star. Having measured
four PK-parameters, we succeeded in performing two independent tests
of general relativity (see Kramer et al. for details) after less than
8 months of observations using PSR~J0737$-$3039A only.  The fifth PK
parameter, the orbital decay, should be determined within 2004.  The
detectability of PSR~J0737$-$3039B makes this system even more
promising, providing further unprecedented tests of gravity theories,
as explained in the contribution of Kramer et al. in this book.

\section{Gravitational Wave Detection}
The discovery of a binary system with the characteristics of
J0737$-$3039 implies a significant increase of the estimates of the
double neutron stars (DNSs) Galactic coalescence rate $\cal{R}$
(Burgay et al. 2003, Kalogera et al. 2004) and, in turn, of the
gravitational waves detection rate for ground based observatories such
as LIGO, GEO and VIRGO.

PSR~J0737$-$3039A and B will coalesce due to the emission of
gravitational waves in a merger time $\tau_{\rm{m}} \approx 85$ Myr, a
timescale that is a factor 3.5 shorter than that for PSR~B1913$+$16
(Taylor, Fowler \& McCulloch 1979). In addition, the estimated
distance for J0737$-$3039 system ($\sim 600$ pc with an intrinsic
uncertainty of about 50\% from the dispersion measure, $\sim 1$ kpc
from X-ray absorption) is an order of magnitude less than that of
PSR~B1913$+$16. These properties have a substantial effect on the
prediction of the rate of merging events in the Galaxy.

For a given class $k$ of binary pulsars in the Galaxy, in fact, apart
from a beaming correction factor, the merger rate $\cal{R}$$_{\rm{k}}$
is calculated (Kim, Kalogera \& Lorimer 2003) as $\cal{R}$$_{\rm{k}}
\propto N_{\rm{k}}/\tau_{\rm{k}}$. Here $\tau_{\rm{k}}$ is the binary
pulsar lifetime defined as the sum of the time since birth,
$\tau_{\rm{b}}$, and the remaining time before coalescence,
$\tau_{\rm{m}}$, whereas $N_{\rm{k}}$ is the scaling factor defined as
the number of binaries in the Galaxy belonging to the given class. The
value of $\tau_b$ for PSR~J0737-3039A can be computed as the time
since the pulsar left the spin-up line (as calculated by Arzoumanian,
Cordes \& Wasserman, 1999) and it results $\sim 150$
Myr. Alternatively one can assume that the characteristic age of
pulsar B is a reliable estimate of the true age of both pulsars. In
this case $\tau_b \sim 50$ Myr. In either cases the lifetime of
PSR~J0737$-$3039 is much shorter than that of PSR~B1913+16
($\tau_{1913}/\tau_{0737} = (365 {\rm{~Myr}})/(235 - 135 {\rm{~Myr}})
\approx 1.6 - 2.7$, where the subscript numbers refer to the pulsars),
implying roughly a doubling of the ratio
$\cal{R}$$_{0737}/\cal{R}$$_{1913}$. 

A much more substantial contribution to the increase factor comes from
the comparison of the pulsars' luminosities. Assuming conservatively a
distance of 1 kpc (as suggested by the X-ray observations; McLaughlin
et al. 2004), the pulsed luminosity at 400 MHz of J0737-3039 binary
system results $L_{0737}\sim 30$ mJy kpc$^2$, much lower than that of
PSR~B1913$+$16 ($\sim 200$ mJy kpc$^2$).  For a planar homogeneous
distribution of pulsars in the Galaxy, the ratio $N_{0737}/N_{1913}$
scales as $L_{1913}/L_{0737} \approx 6$. Therefore we obtain
$\cal{R}$$_{0737}/\cal{R}$$_{1913} \approx 12$. Including the moderate
contribution of the longer-lived PSR~B1534+12 system to the total
rate (van den Heuvel \& Lorimer 1996, Arzoumanian, Cordes \& Wasserman
1999, Kalogera et al. 2001, Kim, Kalogera \& Lorimer 2003), we obtain
an increase factor for the total merger rate
$(\cal{R}$$_{0737}+\cal{R}$$_{1913}+\cal{R}$$_{1534})/(\cal{R}$$_{1913}+\cal{R}$$_{1534})$ of about an order of
magnitude. Similar results have been obtained by Kim et al. (these
proceedings) with a different approach.

This means that ground based gravitational wave detectors such as
LIGO, VIRGO or GEO should be able to detect a burst of gravitational
waves produced in a DNS merger event once every few years instead than
once in few decades, with important consequences for the gravitational
wave community.

\end{document}